\documentclass[letter,aps,twocolumn,groupedaddress]{revtex4-1}
\usepackage[dvips]{graphicx}

\begin{document}
\title{Exact Nonmagnetic Ground State and Residual Entropy of 
$S=1/2$ Heisenberg Diamond Spin Lattices}
\author{Katsuhiro Morita}
\email[e-mail:]{morita@cmpt.phys.tohoku.ac.jp}
\author{Naokazu Shibata}
\affiliation{Department of Physics, Tohoku University, Aoba-ku, Sendai 980-8578 Japan}

\date{\today}
\begin{abstract}
Exactly solvable frustrated quantum spin models consisting of 
a diamond unit structure are presented.
The ground states are characterized by tetramer-dimer states with a
macroscopic degeneracy in a certain range of isotropic exchange 
interaction. The lower bound of the excitation gap is exactly 
calculated to be finite, and the bulk entropy in the limit of 
zero temperature remains finite depending on the shape of 
the boundary of the system. Residual entropy is
in the range of 0 -- 6.1\% of the entropy at high temperature 
for a hexagonal diamond lattice and 0 -- 8.4\% for a square 
diamond lattice. 
These diamond spin lattices are generalized to any dimensions 
and some of them are likely to be synthesized experimentally.
\end{abstract}
\maketitle
Quantum spin systems are fundamental theoretical models of 
interacting electrons, and much effort has
been devoted to understanding the properties of the system for several decades.
Even though their interactions are simple, 
the quantum nature of the spins sometimes leads to unexpected 
behavior at low temperatures. 
The quantum spin liquid (QSL) proposed by Anderson in 1973 
\cite{Anderson} is an example of such a state widely considered to 
be realized in frustrated systems such as the Kagome 
lattice\cite{Elser,Sachdev,QSL}. The QSL state is generally understood as a 
resonating valence bond (RVB) state\cite{Anderson} where all
electrons forming singlet dimers with each other are
dynamically distributed in a certain range of the system without 
any static orders.
Although extensive studies have been directed to find such a QSL state,
the exact ground state of RVB has not yet been obtained
in two- and three-dimensional Heisenberg spin systems. 

One of the few exactly solvable quantum spin models in 2D is the Kitaev model
\cite{Kitaev}.
The interaction of this model, however, has strong anisotropy 
and it is not a general model of magnetic materials
forming an isotropic spin singlet. 
If we restrict the interaction to isotropic exchange coupling,
the Shastry--Sutherland model is known to have an exact solution.\cite{Shastry}
In this model, the ground state is not QSL but a
valence bond solid (VBS) with static dimer order.\cite{Shastry,Shastry2}. 

\begin{figure}
\begin{center}
\includegraphics[width=75mm]{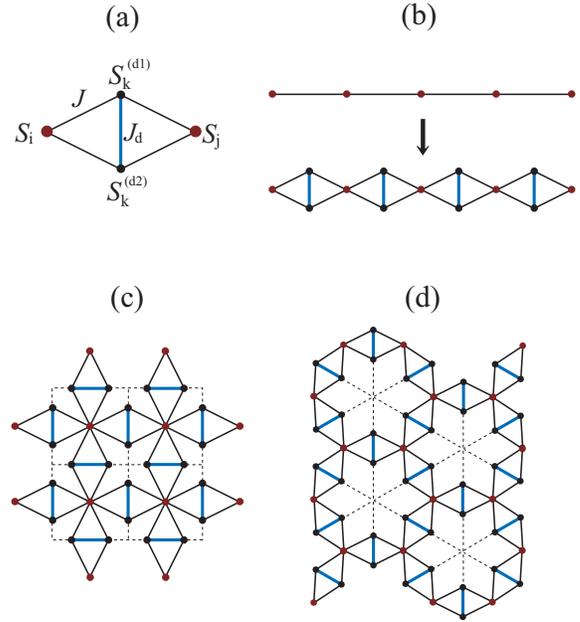}
\caption{(Color online). (a) Diamond unit structure. 
(b) Conversion from Heisenberg chain to diamond chain. 
(c) Square  diamond lattice\cite{DiamondIH}. 
(d) Hexagonal diamond lattice\cite{DiamondIH}. \label{Fig1}} 
\end{center}
\end{figure}

As naively expected, if any configurations of the 
dimers on a lattice have the same energy and they contribute 
to the ground state with equal weight, it would be a similar state 
to the RVB state although it is not a unique ground state.
In this Letter, we present models whose ground state is 
equivalent to an arbitrary configuration of dimers completely 
covering the lattice. 
The model is composed of a diamond unit structure consisting of 
one dimer $\mathbf{S}^{\rm(d1)}_k$-$\mathbf{S}^{\rm(d2)}_k$
and two monomers $\mathbf{S}_i$ and $\mathbf{S}_j$
as shown in Fig.~1(a).
The exactly solvable model is obtained by replacing 
the nearest-neighbor bonds on a regular lattice
with the diamond unit structure provided that
the original lattice can be 
completely covered by dimers.
With this replacement the Heisenberg chain is 
mapped onto the diamond chain\cite{Diamond}, and
square and hexagonal lattices are converted
to square diamond and hexagonal diamond lattices 
as shown in Figs.~1(b) - 1(d).

The Hamiltonian of the models is then written as
\begin{eqnarray} 
H &=& J\sum_{<i,k>} \mathbf{S}_i \cdot( \mathbf{S}^{\rm(d1)}_k +\mathbf{S}^{\rm(d2)}_k ) 
+ J_{\rm d}\sum_k \mathbf{S}^{\rm(d1)}_k\cdot \mathbf{S}^{\rm(d2)}_k \nonumber \\
 &=& J\sum_{<i,k>} \mathbf{S}_i \cdot \mathbf{T}_k  + J_{\rm d}\sum_k \left( \frac{|\mathbf{T}_k|^2}{2} -S(S+1) \right), \label{H}
\end{eqnarray}
where $J$ and $J_{\rm d}$ are the dimer-monomer interaction and dimer intraction, 
respectively.
In this paper, we focus on the case of quantum spins with $S=1/2$.
$\mathbf{S}_i$ and $\mathbf{S}^{({\rm d}j)}_k$ represent monomer and 
dimer spins of $S=1/2$, respectively, 
and $\mathbf{T}_k$ represents the total spin of the dimer,
$\mathbf{T}_k=\mathbf{S}^{\rm(d1)}_k+\mathbf{S}^{\rm(d2)}_k$. 
Since $[\,H,\,{\mathbf T}^2_k\,]=0$ and 
${\mathbf T}^2_k=T_k(T_k+1)$, each eigenstate of $H$ is characterized by 
a set of $\{T_k\}$. 
In the following, we show that the ground state of the Hamiltonian 
is exactly solvable when $J_{\rm d}>J>0$. 

\begin{figure}
\begin{center}
\includegraphics[width=60mm]{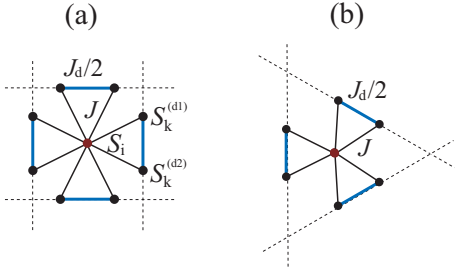}
\caption{(Color online). Interactions included in a
partial Hamiltonian $h_i$ in the diamond spin lattice.         
(a) One unit of square diamond lattice ($n_{\rm d}\!=\!4$). 
(b) One unit of hexagonal diamond lattice ($n_{\rm d}\!=\!3$). 
\label{Fig2}}
\end{center}
\end{figure}

We first decompose $H$ to the sum of partial Hamiltonians $h_i$ as
\begin{equation}
H = \sum_{i}^N h_i,
\end{equation}
where $N$ is the number of units in the diamond spin lattice that encloses 
one monomer spin as shown in Figs.~1(c) and 1(d) by the broken lines.
Then, the minimum energy $E_{\rm (min)}$ of $H$ and 
the sum of the minimum energy $e_{i ({\rm min})}$ of $h_i$
satisfy the inequality
\begin{equation}
E_{\rm (min)} \ge \sum_{i}^N e_{i {\rm (min)}} \label{ineq}
\end{equation}
for any choice of $h_i$ based on the variational principle.
The equality occurs only when the eigenstate of $H$ is also the 
lowest energy eigenstate of each $h_i$.
Here, we chose $h_i$ as one unit of the lattice shown in Fig.~2, 
where the exchange interaction between the two spins 
$\mathbf{S}^{\rm(d1)}_k$ and $\mathbf{S}^{\rm(d2)}_k$ is written 
as ${J_{\rm d}/2}$ 
because of the double counting of the interaction at the interface
between the two neighboring units,
\begin{equation}
h_i  = J\sum_{k}^{n_{\rm d}} \mathbf{S}_i \cdot \mathbf{T}_k  + \frac{J_{\rm d}}{2}\sum_k^{n_{\rm d}} \left( \frac{|\mathbf{T}_k|^2}{2} -S(S+1) \right) .
\end{equation}
We first consider the eigenstates of the partial Hamiltonian $h_i$, where 
the number of dimers is $n_{\rm d}$.
Since the total spin of the dimer is a conserved quantity, $[\,h_i,\,{\mathbf T}^2_k\,]=0$, 
their sum $n_{\rm t} \equiv \sum_{n_{\rm d}} T_k$ in each unit
is a good quantum number.
This means that both the numbers of triplet dimers, $n_{\rm t}$, and 
singlet dimers, $n_{\rm d} - n_{\rm t}$, are conserved quantities in each unit.
The eigenstates of $h_i$ are then obtained by solving a problem of
interacting $S = 1$ triplet dimer spins and the $S = 1/2$ 
monomer spin at the center of the unit. 
The local Hamiltonian $h_i$ is now reduced to
\begin{equation}
{h_i}  = J \sum_{k=1}^{n_{\rm t}} \mathbf{S}_i \cdot {\cal T}_k
+ \frac{1}{8}J_{\rm d}(4n_{\rm t}-3n_{\rm d}) ,
\end{equation}
where $ \mathbf{S}_i$ is $S=1/2$ spin at the center and 
${\cal T}_k $ is $S=1$ spin representing three states of triplet dimers.
From the Lieb--Mattis theorem\cite{Lieb-Mattis}, the ground state belongs to $S_{\rm tot} =n_{\rm t} - 1/2$
for the antiferromagnetic coupling $J>0$, and is exactly solvable with the energy
${e}_{\rm min} = -J(n_{\rm t}+1)/2$. 
The lowest energy $e_{\rm min}(n_{\rm t})$
of $h_i$ under the condition of fixed $n_{\rm t}$ is 
\begin{equation}
e_{\rm min}(n_{\rm t}) = \left\{
\begin{array}{ll}
- \frac{1}{2}J(n_{\rm t}+1) + \frac{1}{8} J_{\rm d}(4n_{\rm t} - 3n_{\rm d}) & 
(n_{\rm t}\!\ge\!1) \\ 
-\frac{3}{8} J_{\rm d} n_{\rm d} & (n_{\rm t}\!=\!0) .
\end{array}
\right. 
\end{equation}
The lowest energy $e_{\rm min}$ of $h_i$ is then obtained as
\begin{equation}
e_{\rm min} = \left\{
\begin{array}{ll}
-\frac{1}{2}J(n_{\rm d}+1) + \frac{1}{8}J_{\rm d}n_{\rm d} &(J_{\rm d}\!<\!J,\  n_{\rm t}\!=\!n_{\rm d})\\
-J + \frac{1}{8}J_{\rm d}(4-3n_{\rm d}) &(J\!<\!J_{\rm d}\!<\!2J,\  n_{\rm t}\!=\!1)\\
-\frac{3}{8}J_{\rm d}n_{\rm d}  & (2J\!<\!J_{\rm d},\  n_{\rm t}\!=\!0) 
\end{array}
\right.
\end{equation}
depending on the ratio $J_{\rm d}/J$. 
In the region of $J_{\rm d}\!>\!J$, the number of triplet 
dimers in $h_i$ is 0 or 1.
As shown in the following, the ground state and its energy of $H$
are exactly obtained. The ground state of $H$ in the case of 
$n_{\rm t} = 0 $ in  each unit is obvious. 
All the spins of dimers, $\mathbf{S}^{\rm(d1)}_k$ and $\mathbf{S}^{\rm(d2)}_k $,
form spin singlet states and decouple from the other part of the system.
The ground state is then characterized by the dimer-monomer (DM) state. 

\begin{figure}
\begin{center}
\includegraphics[width=60mm]{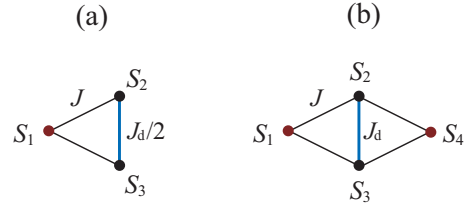}
\caption{(Color online). (a) Triangle structure in $h_i$. (b) Quadrangle structure at
the interface of two adjacent $h_i$.
\label{Fig3}}
\end{center}
\end{figure}

Here, we concentrate on the case of $J\!<\!J_{\rm d}\!<\!2J$
where only one triplet dimer is confined in each unit ($n_{\rm t} = 1$).
Since spin singlet dimers are decoupled from the rest of the system, 
we consider three spins on one triangle in the unit consisting of one
triplet dimer and one monomer spin, as shown in Fig.~3(a). 
The ground states of the three spins are doubly 
degenerate with different $z$-components of the total spin as shown below:   \\
\begin{eqnarray}
|gs\rangle_{\rm 123}^{+} &=&  \frac{1}{\sqrt{3}}\left[\sqrt{2}|\!\downarrow  \, \rangle_1|t\rangle^+_{23} -|\!\uparrow \,  \rangle_1|t\rangle^0_{23} \right], \label{tri+} \\
|gs\rangle_{\rm 123}^{-} &=&  \frac{1}{\sqrt{3}}\left[\sqrt{2}|\!\uparrow  \, \rangle_1|t\rangle^-_{23} -|\!\downarrow \,  \rangle_1|t\rangle^0_{23} \right], \label{tri-}
\end{eqnarray}
where
\begin{eqnarray}
|t\rangle^+_{ij} &=&  |\!\uparrow \, \rangle_i|\!\uparrow \, \rangle_j , \\
 |t\rangle^0_{ij} &=&  \frac{1}{\sqrt{2}}(|\!\uparrow \, \rangle_i|\!\downarrow \, \rangle_j  + |\!\downarrow \, \rangle_i|\!\uparrow \, \rangle_j) , \\
 |t\rangle^-_{ij} &=&  |\!\downarrow \, \rangle_i|\!\downarrow \, \rangle_j .
\end{eqnarray}
The energy of these states is obtained as
\begin{equation}
e_{\rm tri(min)}=-J+\frac{1}{8}J_{\rm d} \label{e_tri}
\end{equation}
with the interaction of $J_{\rm d}/2$ between the two 
spins of the triplet dimer labeled by the numbers 2 and 3 in Fig.~3(a).
We then couple two triangles by joining the edge bonds of $J_{\rm d}/2$ 
to form a quadrangle shown in Fig.~3(b).
The exact ground state $|gs\rangle_{\rm qua}$ and its energy 
$e_{\rm qua(min)}$ are obtained as 
\begin{eqnarray}
|gs\rangle_{\rm qua} =  \frac{1}{\sqrt{3}}\left[|t\rangle^+_{14}|t\rangle^-_{23}+|t\rangle^-_{14}|t\rangle^+_{23} -|t\rangle^0_{14}|t\rangle^0_{23} \right], \label{gs_qua} 
\end{eqnarray}
\begin{equation}
e_{\rm qua(min)}=-2J+\frac{1}{4}J_{\rm d} \label{e_qua},
\end{equation}
where $|gs\rangle_{\rm qua}$ is rewritten using Eqs. (\ref{tri+}) and (\ref{tri-}) as
\begin{eqnarray}
|gs\rangle_{\rm qua}  &=& \frac{1}{\sqrt{2}}(|gs\rangle_{\rm 123}^{+} |\!\downarrow  \, \rangle_4 + |gs\rangle_{\rm 123}^{-}|\!\uparrow  \, \rangle_4) , \nonumber \\
&=& \frac{1}{\sqrt{2}} (|gs\rangle_{\rm 423}^{+} |\!\downarrow  \, \rangle_1 +|gs\rangle_{\rm 423}^{-}|\!\uparrow  \, \rangle_1)  \label{tri423}
\end{eqnarray}
with $|gs\rangle^{+(-)}_{\rm 432}$ being obtained by replacing $\lq\lq 1 \textquotedblright$ with $\lq\lq 4 \textquotedblright$  in Eqs. (\ref{tri+}) and (\ref{tri-}).

Comparing Eqs.~(\ref{e_tri}) and (\ref{e_qua}),
we find that the minimum energy of the quadrangle is 
just twice the minimum energy of the triangle.
This means that the minimum energy of the 
sum of two partial Hamiltonians $h_i+h_j$ sharing 
one triplet dimer state at the interface is equal to twice the 
minimum energy of each partial Hamiltonian. Indeed, we find that
the minimum energy eigenstates of $h_i+h_j$ are
also the minimum energy eigenstates of each $h_i$, as shown in Eq. (\ref{tri423}). 
Since the eigenstates of the quadrangle $|gs\rangle_{\rm qua}$
are decoupled from the rest of the 
system by the singlet dimer states, $|gs\rangle_{\rm qua}$ is also
an eigenstate of $H$.
If $H$ is composed of such coupled 
partial Hamiltonian $h_i+h_j$, the ground state 
is exactly obtained as product states of 
$|gs\rangle_{\rm qua}$ surrounded by spin singlet dimer states.
We refer to this state as the tetramer-dimer (TD) state.
The ground-state energy of $H$ per unit of the diamond spin lattice  
is then given by
\begin{equation}
E_{\rm (min)}/N=-J+\frac{1}{8}(4 -3\bar{n}_{\rm d})J_{\rm d}
\label{E_min},
\end{equation}
where $\bar{n}_{\rm d}= N_{\rm d}/N$ is the ratio of 
the total number of dimers, $N_{\rm d}$, to 
the total number of monomers, $N$, in the diamond spin lattice. 
The condition to compose $H$ from the 
coupled partial Hamiltonian $h_i+h_j$ whose eigenstate
consists of quadrangle $|gs\rangle_{\rm qua}$ is to find
a complete dimer covering the original lattice 
of monomers, as shown in Fig.~4.
The number of dimer configurations 
depends on the shape of the boundary but 
it generally increases to a macroscopic number with the 
increase in the system size yielding finite residual 
entropy in the ground state as shown later.
Note that even in the case of $J_{\rm d} < J$, where
equality in Eq.~(\ref{ineq}) does not hold, 
there is a possibility that the TD state is still the ground state. 
Equation~(\ref{ineq}) is the inequality determining the lower limit 
of the total Hamiltonian, and thus $J_{\rm d}= J$ is the upper bound 
of the transition to the TD state.
Indeed, the diamond chain has the transition point from
Ferri to TD state at $J_{\rm d} \sim 0.909$\cite{Diamond}.

\begin{figure}
\begin{center}
\includegraphics[width=52mm]{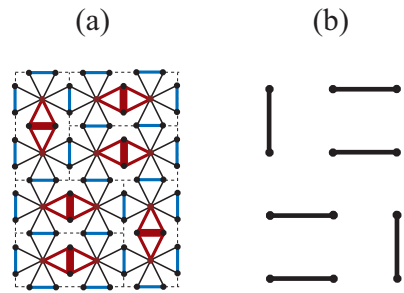}
\caption{(Color online). (a) TD state on square diamond lattice. Red thick lines 
represent spin triplet dimers and blue thin lines represent spin singlet 
dimers. (b) Corresponding dimer configuration on the original lattice of
monomers. 
\label{Fig4}}
\end{center}
\end{figure}

We next show that only TD states have the 
lowest energy in the region of $J < J_{\rm d} < 2J$.
Each eigenstate of $H$ is characterized by 
a set of $\{T_k\}$ and there is the inequality Eq. (\ref{ineq}) 
requiring that each $h_i$ in $H$ has the lowest energy 
with one triplet dimer ($n_{\rm t}=1$).
Since each triplet dimer is shared by the neighboring two 
units of the diamond spin lattice, the minimum energy state has triplet dimers 
whose number $N_{\rm t}$ in the total system is half 
the number of units, $N_{\rm t}=N/2$. 
This condition is equivalent to the complete dimer covering 
and the definition of the TD state. 
The uniqueness of the ground state in the 
subspace of $N_{\rm t}=N/2$ and $n_{\rm t}=1$ in each unit 
is shown by exactly solving a tetramer of 4 spins.
Since the condition of $N_{\rm t}=N/2$ and $n_{\rm t}=1$ in each unit 
means that the spin singlet dimers completely surround
the tetramer of 4 spins, all the tetramers in the system are
independent of each other.
The exact diagonalization of the tetramer shows that
the ground state is singlet and separated from the 
lowest excited states by the energy gap $\Delta E=J$,
which shows that only the TD states are the lowest energy states.

\begin{figure}
\begin{center}
\includegraphics[width=80mm]{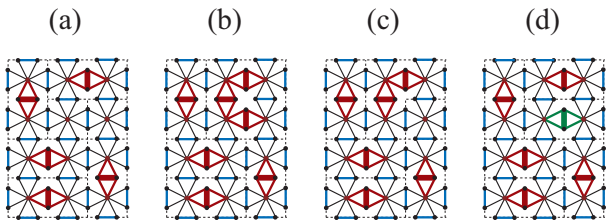}
\caption{(Color online). Elemental excitations from TD state.
(a) $N_{\rm t}=N/2-1$. (b) $N_{\rm t}\ge N/2+1$.  
(c) $N_{\rm t}=N/2$, $n_{\rm t}$ = 0 and 2 at one pair of next-nearest-neighbor units and $n_{\rm t} = 1$ in other units.
(d) $N_{\rm t}=N/2$ and $n_{\rm t}=1$ in each unit with the excitation in 
one tetramer (green lines).
\label{Fig5}}
\end{center}
\end{figure}

Since $[\,H,\,{\mathbf T}^2_k\,]=0$ and $[\,h_i,\,{\mathbf T}^2_k\,]=0$, the lower bound of the energy of any subspace of \{$T_k$\} is obtained by the variational principle represented by Eq. (3).
The elemental excitations from the TD states are then 
classified by the quantum numbers $N_{\rm t}$ and $n_{\rm t}$ as
follows:
\begin{itemize}
\item{$N_{\rm t}=N/2-1$ that is obtained by replacing one 
tetramer with one spin singlet dimer and two monomers, as in Fig.~5(a).
The lowest excitation energy is obtained as $\Delta E_1=2J-J_{\rm d}$.}
\item{$N_{\rm t}\ge N/2+1$ that is obtained by replacing at least one 
spin singlet dimer with a spin triplet dimer, as in Fig.~5(b).
Since $J_{\rm d} = J$ is the upper bound of the TD phase boundary,
the excitation energy has a lower bound as $\Delta E_2 \ge J_{\rm d} - J$.} 
\item{$N_{\rm t}=N/2$, $n_{\rm t}$ = 0 and 2 at one pair of next-nearest-neighbor units and $n_{\rm t} = 1$ in other units 
that is obtained by exchanging one tetramer and an adjacent spin singlet 
dimer, as in Fig.~5(c).
The lowest excitation energy is given as $\Delta E_3=0.6185J$ by the exact diagonalization.
Equations (3) and (6) show that other configurations of $N_{\rm t}=N/2$ and $n_{\rm t}\neq 1$ have a lower bound of excitation energy, $\Delta(3) \ge J$ .} 
\item{$N_{\rm t}=N/2$ and $n_{\rm t}=1$ in each unit that is the excitation within 
a tetramer in a TD state as in Fig.~5(d).
The excitation energy is obtained as $\Delta E_4=J$. }
\end{itemize}
We therefore conclude that in the region of $J<J_{\rm d}<2J$,
only the TD state is the ground state and a
finite excitation gap separates the TD ground state from the 
other excited state.
The lowest excitation energy $\Delta E_{\rm min}$ 
is then summarized as
\begin{center}
\begin{math}
\begin{array}{ll}
\Delta E_{\rm min} \ge J_{\rm d} - J, & (J < J_{\rm d} < \frac{3}{2}J), \\
\Delta E_{\rm min} = 2J-J_{\rm d} . & ( \frac{3}{2}J \le J_{\rm d} < 2J).
\end{array}
\end{math}
\end{center}

All the above results on the ground state and excitation energy
are common features of the diamond spin lattice 
including one- and three-dimensional systems and any boundary conditions provided that TD states are  constructed on the lattice.
Experimentally, it would be relatively easy to 
synthesize 2D hexagonal or square diamond lattices, 
because the lattice structure is equivalent
even if dimers are disposed perpendicular to the 2D plane.
There are several reports on the bimetallic polymeric coordination compounds
that have a hexagonal diamond lattice structure\cite{Diamond2,Diamond3,Diamond4} 
and a square diamond lattice structure\cite{Diamond5,Diamond6}. 
We do not discuss the case of $J_{\rm d} < J$,
because results are model-dependent.

We finally comment on the residual entropy of the TD ground state.
The number of degenerate TD ground states
is equivalent to the number of configurations of
dimer covering the original lattice of the monomer spin.
This is known as dimer problems and the number of dimer 
configurations, $N_{\rm g}$, is given for an $m\times n$ square 
lattice\cite{Fisher,Kasteleyn} with torus and open boundary conditions as 
\begin{equation}
\lim_{N\rightarrow\infty} \frac{1}{N}\ln{N_{\rm g}}=0.2915609,\ \ \ (N=nm)
\end{equation}
which corresponds to the residual bulk entropy of the ground state
per unit of diamond spin lattice. 
For a hexagonal lattice under torus boundary conditions, 
it is shown that\cite{Wannier,Kasteleyn2}
\begin{equation}
 \lim_{N\rightarrow\infty}\frac{1}{N}\ln{N_{\rm g}}=0.169157
\end{equation}
in the bulk limit.
However, it is also known that $N_{\rm g}$
depends on the shape of the boundary of the lattice even in the bulk limit.
For example, $\lim_{N\rightarrow\infty}\frac{1}{N}\ln{N_{\rm g}}=$ 0 -- 0.130812 for 
a hexagonal lattice with open boundary conditions\cite{Elser2}, and 
$\lim_{N\rightarrow\infty}\frac{1}{N}\ln{N_{\rm g}}=0$ for a square 
lattice of almost square-shaped open boundaries\cite{Sachs}.
 Since the entropy under the torus boundary condition is expected to be the largest, the residual entropy of the TD ground state is distributed
at least in the range of 0 -- 6.1\% of the entropy at high temperature 
for a hexagonal diamond lattice and 0 -- 8.4\% for a square 
diamond lattice depending on the boundary conditions. 
These results show that the residual entropy of 
diamond spin lattices is boundary-dependent even in the 
bulk limit, which may lead to quite unusual
thermodynamic behavior at low temperatures.
If we include next-nearest-neighbor exchange interactions in a square diamond lattice, 
the second-order perturbation analysis shows that the low-energy effective model will be a quantum dimer model\cite{qdm} that releases the residual entropy and may lead to a quantum spin liquid state.

\section*{Acknowledgment} 
The present work was supported by a Grant-in-Aid for
Scientific Research (No. 26400344) from JSPS.


\begin{thebibliography}{99}
\bibitem{DiamondIH} The Ising-Heisenberg model on the same diamond spin lattice is studied in Refs. 10 and 11.
\end{thebibliography}
\end{document}